\documentclass[traditabstract]{aa}
\usepackage{txfonts}
\usepackage{graphicx}
\usepackage{epsfig}

\begin{document}

\title{The model constraints from the observed trends for the \\ quasi-periodic
oscillation in RE J1034+396}

\author{B. Czerny\inst{1}\and P. Lachowicz\inst{2} \and M. Dov\v{c}iak\inst{3}
\and V. Karas\inst{3}  \and T. Pech\'{a}\v{c}ek\inst{3} \and Tapas K. Das\inst{4}}

\institute{Nicolaus Copernicus Astronomical Center, Polish Academy of Sciences,
ul.~Bartycka~18, 00-716~Warszawa, Poland,\\ \email{bcz@camk.edu.pl}
\and
Centre for Wavelets, Approximation and Information Processing, Temasek
Laboratories at National University of Singapore,\\ 5A Engineering Drive~1,
\#09-02~Singapore~117411, \email{pawel@ieee.org}
\and
Astronomical Institute, Academy of Sciences, Bo\v{c}n\'{\i}~II~1401,
CZ-14131~Prague, Czech~Republic
\and
Harish Chandra Research Institute, Allahabad 211019, India}

\date{Received; Accepted}

\abstract{We analyze the time variability of the X-ray emission of 
RE~J1034+396 -- an active galactic nucleus with the first firm detection of
a quasi-periodic oscillations (QPO). Based on the results of a wavelet
analysis, we find a drift in the QPO central frequency. The change in the QPO 
frequency correlates with the change in the X-ray flux with a short time delay.
The data specifically suggest a linear dependence between the QPO period and
the flux, and this gives important constraints on the QPO models. In particular, 
it excludes explanation in terms of the orbiting hot spot model close to a black hole.
Linear structures such as shocks, spiral waves, or very distant flares are favored.}

\authorrunning{Czerny et al.}
\titlerunning{The model constraints for the QPO in RE J1034+396}

\keywords{Accretion, accretion disks -- Galaxies: active -- Galaxies:
individual: RE J1034+396}

\maketitle

\section{Introduction}

Accretion onto black holes is a non-stationary process. Strong variability is
seen in the X-ray emission of both Active Galactic Nuclei (AGN) and Galactic
Black Holes (GBH). This variability is mostly aperiodic, but quasi-periodic
modulations occasionally appear (van der Klis 1989). The quasi-periodic oscillation (QPO) phenomenon is well
known in GBH where the frequency can be as high as the orbital frequency in the
inner disk. Recently the QPO appearance has been reported to occur in a bright
Seyfert~1 galaxy, RE J1034+396 (Gierli\'{n}ski et al. 2008; hereafter G08)
and BL~Lacertae object PKS~2155$-$304 (Lachowicz et al. 2009). The significance 
of the signal in RE J1034+396 
($f_0\simeq 2.6 \times 10^{-4}$~Hz, period $P_0\simeq 3733$~s) is very high, 
($>3\sigma$) as determined by G08 using Vaughan's
(2005) test. The result was recently confirmed by Vaughan (2010) with a Bayesian approach 
(although with slightly lower confidence). The existence of this QPO was used to constrain
the properties of the warm absorber in this source (Maitra \& Miller 2010).

Despite a number of scenarios proposed in the literature, neither the origin of
stochastic variability nor the QPO phenomenon have been well understood (for
reviews see e.g. van der Klis 2006; McClintock \& Remillard 2006; Done et al.
2007). The standard variability studies rely on the Fourier analysis and the
discussion of the time-averaged power spectrum (Feigelson \& Babu 1992). 
For the GBH the time dependence of the QPO frequency has also been studied, but the time
bins were always set long in comparison to the disk Keplerian timescale because
the latter is on the order of milliseconds for GBH. In AGN the Keplerian
time-scale is by a factor of $\sim 10^6$ longer, so the discovery of QPOs in
AGNs opens new possibilities for the investigation of accreting black holes.

Similar to Fourier spectra methods, the wavelet analysis  provides fast linear 
operations on data vectors. Unlike the Fourier transform, the base functions of wavelets
are localized, and this brings new opportunities to the data analysis.

The wavelet analysis can be particularly useful in the exploration of the QPO
phenomenon (e.g. Scargle et al. 1993; Lachowicz \& Czerny 2005; Espaillat
et al. 2008; Lachowicz \& Done 2010). In this paper we 
construct for the the first time a wavelet map for RE~J1034+396 data when the
source showed the QPO activity. We find the QPO period to change in a
time-frequency (or, equivalently, time-period) plane and we 
investigate the properties of this pattern.
Based on the wavelet results we discuss possible constraints on the physical
origin of the QPO in RE~J1034+396.

\section{Wavelet analysis of RE~J1034+396 light curve}

\begin{figure}
\begin{flushright}
\includegraphics[angle=0,width=0.4585\textwidth]{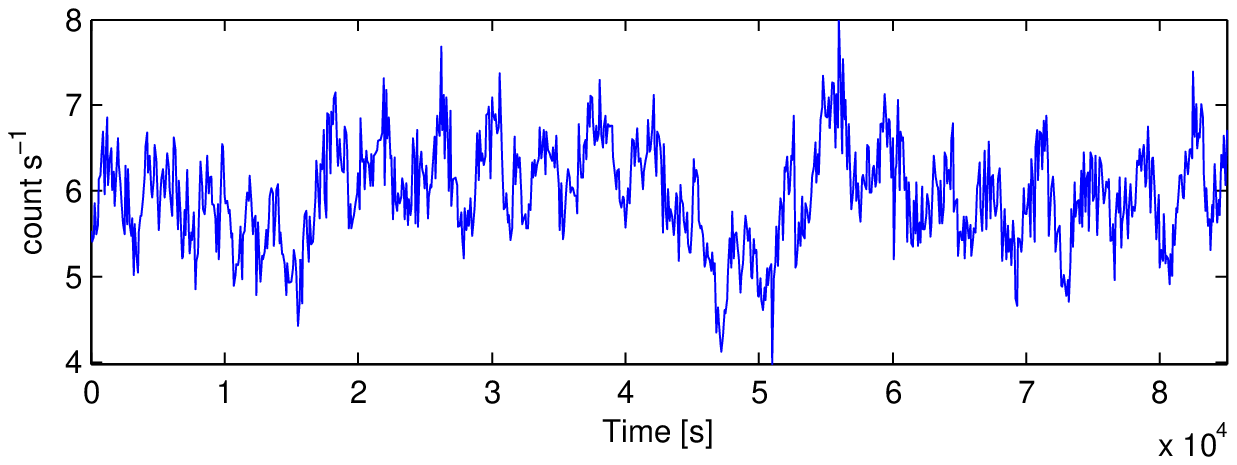}
\includegraphics[angle=0,width=0.475\textwidth]{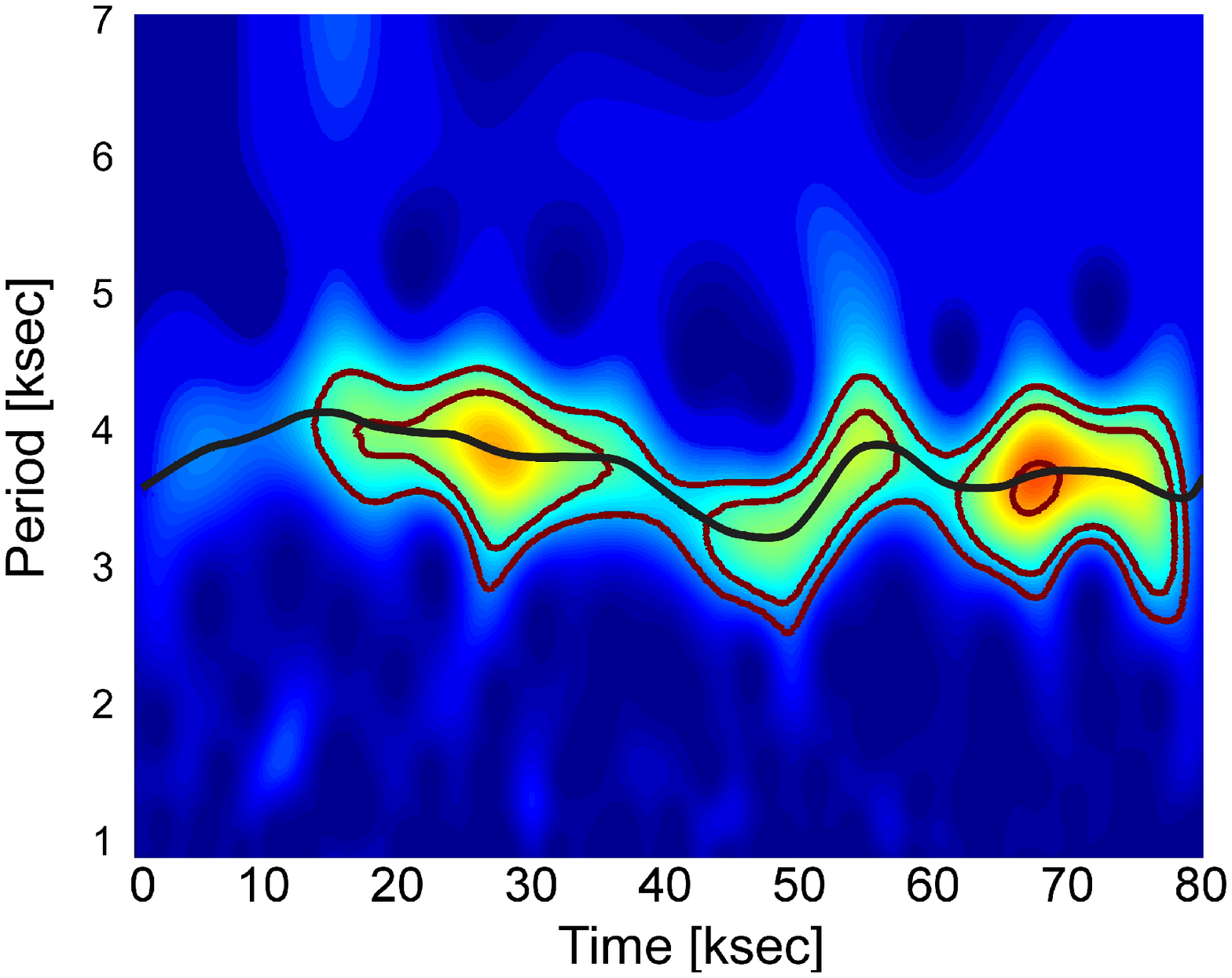}
\end{flushright}
\caption{{\it (upper)} X-ray light curve of RE~J1034+396, and {\it (bottom)}
the corresponding wavelet map. The increasing values of wavelet power are
denoted by gradual changes of the colours, i.e. from deep blue, through
green and yellow to red. The confidence contours of 90\%, 95\%, and 99\% are marked, and they are shifted with respect to the color levels (see text for the discussion). A solid black line traces the the QPO period.}
\label{fig:map}
\end{figure}

The source RE J1034+396 was observed by {\it XMM-Newton} satellite on 2007~May~31, 
and the resulting X-ray light curve analysis was reported in
G08. Using the same procedure, we extract the source light curve in the
0.3$-$10~keV energy band, which covers $8.5\times 10^4$~s of continuous
observations sampled evenly every $\Delta t\!=\!100$~s. We use the
wavelet analysis  codes of Torrence \& Compo (1998) 
similarly as we did in Lachowicz \& Czerny
(2005). Morlet wavelet is adopted, with the standard assumption $2 \pi \nu_o = 6$.
We limit the scale range to the region of the expected QPO signal.
The result is presented in Fig.\ref{fig:map} together with corresponding
light curve. 

The wavelet map confirms the presence of the QPO in the source at
the period $P_0\simeq 3733$~s through  a range of peaks. In order to assign the 
confidence levels to those peaks we have performed Monte Carlo simulations. We 
adopted the underlying broadband power spectrum in the form of a broken power law, with the
normalization and the high-frequency slope 1.35 determined by G08 for this observation of 
the source, and we
assumed the frequency break at $10^{-5}$ Hz, in consistency with the data. 

The simulated lightcurves
were created using the algorithm of Timmer \& K{\"o}nig 1995, of the duration and timestep 
consistent with the observed lightcurve. We created 7500 lightcurves and the corresponding 
wavelet maps, and built the statistics for each of the scales in the discussed range. This allowed us
to assign significance levels for each wavelet scale independently, which is important in the
lightcurves with underlying red noise background. The average wavelet values rise toward longer 
periods, as does the power density spectrum, and this drift causes the observed misalignment between the
colors on the map (referring to absolute values) and confidence levels in Fig.~\ref{fig:map}.

The significant regions are within a period range of 3000 - 4500 s, but they 
form a complex pattern. We see some very high amplitude peaks as well 
as indications of the period change. The QPO
signal is relatively weak at the beginning of the data, as noticed by G08, 
who divided the lightcurve into two parts: Segment 1 to $2 \times 10^4 $ s and 
Segment 2 above that time. In Segment 2 the QPO detection is almost always above the 90 percent 
confidence.

We thus investigate the wavelet map in greater detail. For every time
step of $\Delta t$ we fit the intensity peaks in the wavelet map with a Gaussian
profile that allows us to determine the best fit of the QPO
period and the wavelet amplitude at the peak. We denote this line with a 
black solid line in the wavelet map
(see Fig.~\ref{fig:map}, lower panel). The errors of the peak position were estimated
at the basis of the largest error in the flux determination (0.26 cts s$^{-1}$). The peak position 
errors are  large in the Segment 1, but they are well constrained in Segment 2.

The QPO period marked by the peak positions seems to follow a
complex  pattern in the time-period plane. The changes in the frequency are significant:
the $\chi^2$ fit to this pattern assuming constant period value gave unacceptable results 
(reduced $\chi^2$ = 6.9).

\begin{figure}
\includegraphics[angle=0,width=0.475\textwidth]{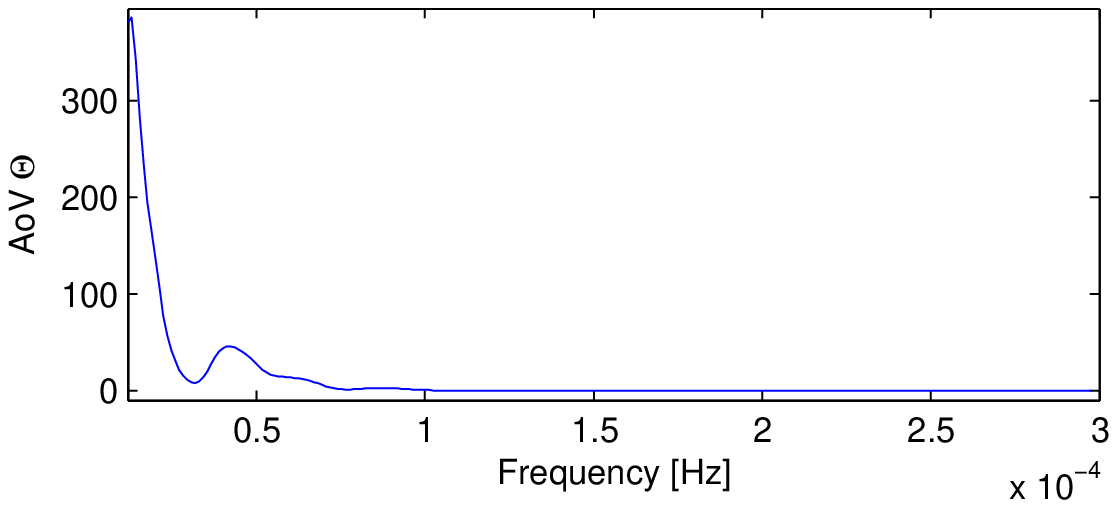}
\includegraphics[angle=0,width=0.475\textwidth]{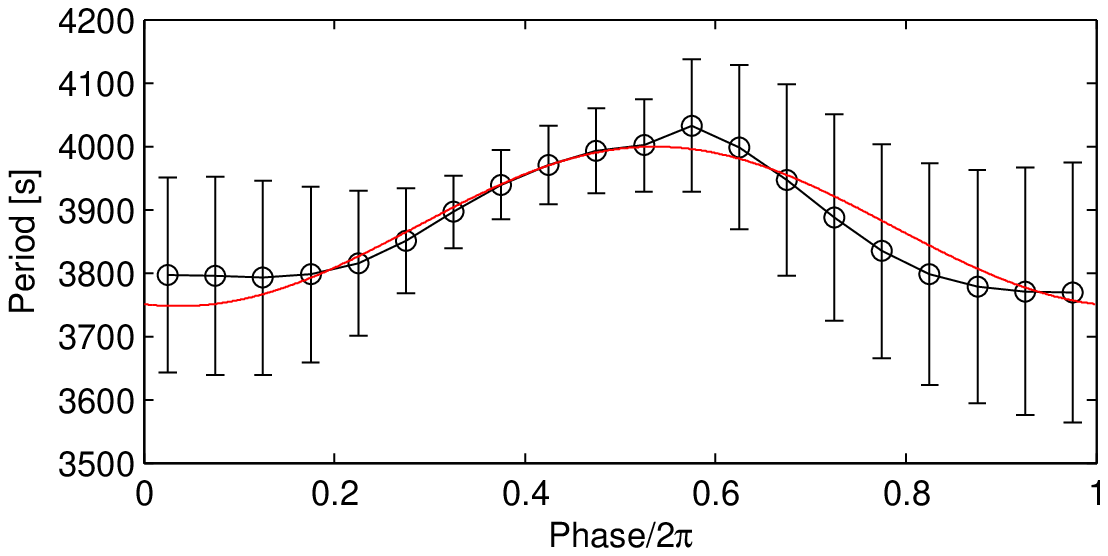}
\caption{{\it (upper)} AoV periodogram for a QPO period indicating
additional periodicity at 24008 s. {\it (bottom)} The QPO
period curve folded with a period of $24008$~s. Although the presence of that
additional period  would be interesting, its formal significance is very low.}
\label{fig:aov}
\end{figure}
\begin{figure}
\includegraphics[angle=0,width=0.475\textwidth]{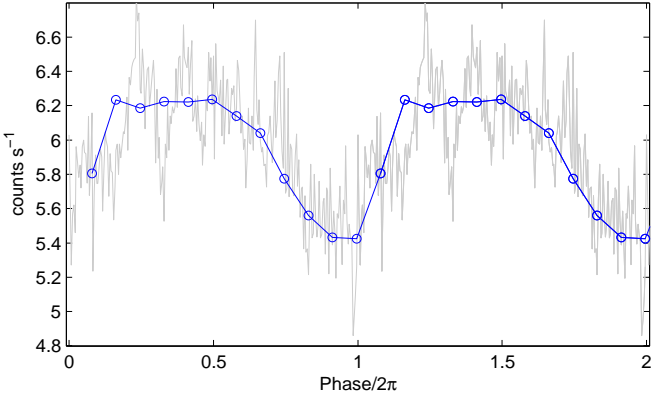}
\caption{Folded light curve with a period of $24$~ks (blue line), derived
for the second part of the light curve from Fig.~\ref{fig:map}, i.e from 
$2.5 \times 10^{4}$ s till the end of the data (gray line); plotted over two
periods for clarity.}
\label{fig:folded_curve}
\end{figure}

In order to find out whether any additional periodicity is
present in the system, we analyzed a temporal variability of the QPO period. We
employ the Analysis of Variance (AoV) method of Schwarzenberg-Czerny (1996) and
calculate the AoV periodogram for the QPO period curve. We allow for an 
oversampling by a factor of 5 in the periodogram because otherwise the
resolution is  insufficient for a tentative period detection at the
lowest frequencies. The periodogram analysis indicates a time-scale 
of $\sim$ 24008~s (see Fig.~\ref{fig:aov}, upper panel).  After Lachowicz et
al. (2006) we determine its significance, $P_1=0.5982$, to be low, i.e.
at the 40.18\% of confidence ($\Theta\!=\!46.1$, $n\!=\!851$, $n_{\rm
corr}\!=\!70.91$, $N\!=\!1$). The actual value may be even lower because the data
points obtained from the wavelet map are partially correlated (Maraun \& Kurths
2004).
The folded QPO period curve is shown in bottom panel of Fig.~\ref{fig:aov}. It
displays approximately a sinusoidal trend, but with very large errors.  We also 
checked for the presence of this putative period directly in the X-ray lightcurve. 
The folded X-ray light curve (starting at
$2.5\times 10^4$~s; Segment 2 in notation of G08) with the 24~ks period is shown in 
Fig.~\ref{fig:folded_curve}).
Again, a pattern seems to be present there, but the formal significance of the
24008 s periodicity in the X-ray light curve estimated using the Vaughan (2005) test
is low because of the fairly high red-noise level.

Concluding, the temporal modulation of the QPO period on the 24~ks
time-scale is visible, both the period and the wavelet amplitude vary in time,
and the X-ray light curve follows the pattern but the modulation cannot be
considered as firmly periodic for the present data. However, this
outcome suggests that some underlying mechanism must be present
 and responsible for observed modulation of the QPO amplitude, the
frequency and the flux. Below, we study a correlation between X-ray flux
and QPO period without referring to any periodicity in their correlated
behaviour.

\section{Flux versus QPO period trend}

\begin{figure}
\includegraphics[angle=0,width=0.475\textwidth]{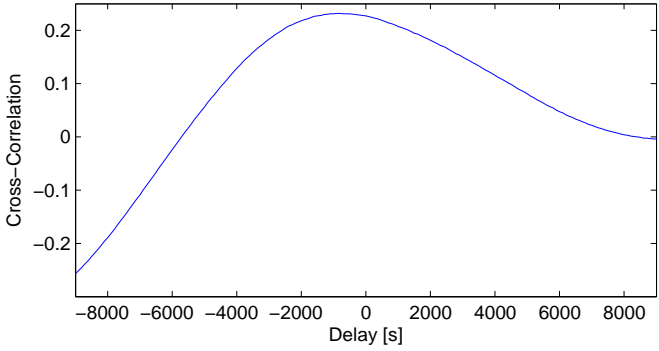}
\caption{Normalized cross-correlation function of 
the QPO period and the X-ray light curve over the whole data range. The 
flux change is delayed by $\sim 900$~s with respect to the change in the
QPO period.}
\label{fig:ccf}
\end{figure}

The wavelet map shows a strong decrease in the marked QPO period at the time
about $5 \times 10^4$ s, which is accompanied by a decrease in the X-ray flux.
This motivated us to check for the overall correlation between the flux and the 
QPO period in the data.
The cross-correlation function between the X-ray luminosity and the QPO
period is shown in Fig.~\ref{fig:ccf}. The correlation is present, and the 
time delay between the curves is $\sim$900~s, with the flux lagging behind the
frequency change. The delay is by a factor 4.1 shorter than the QPO period itself. 
If only the second part of the data is investigated (Segment
2 in notation of G08), the measured delay is somewhat longer ($1400$~s).

\begin{figure}
\includegraphics[angle=0,width=0.475\textwidth]{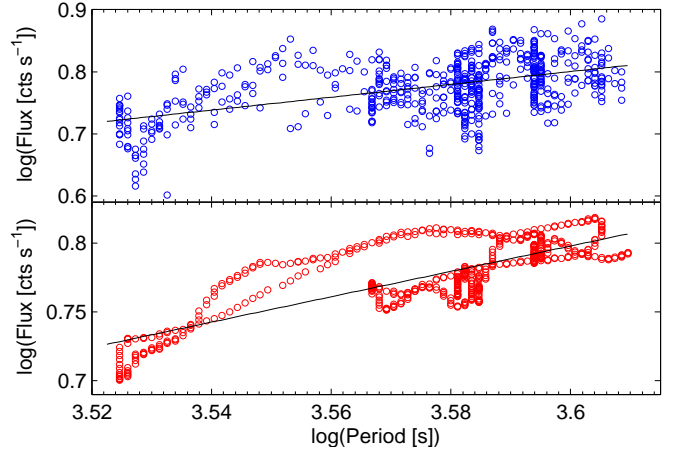}
\caption{{\it (upper)} Flux vs. QPO period relation in the Segment 2 of the
data for original X-ray light curve, and {\it (bottom)} the smoothed light
curve.}
\label{fig:flux_period}
\end{figure}

Therefore, we made a logarithmic plot (see Fig.~\ref{fig:flux_period}, upper
panel) of the dependence of the flux on the observed QPO period allowing for a
shift according to the measured delay. If the full light curve is used, with the
delay of 900 s, the correlation coefficient $r$ between the quantities is very
low, 0.26, and the slope of the best-fit straight line 
is $0.42 \pm 0.05$. If only the Segment~2 is used with the delay of 1400 s, $r =
0.52$, and the slope is $1.03 \pm 0.07$. This means that the correlation is much
more significant, and the slope is close to 1. 

We justified the omission of Segment 1 by examining the flux--period relation for the QPO in this part of the data. Visually, there seems to be a dip in the flux at time $\sim 1.6 \times 10^4$ s accompanied by a rise rather than a fall in period, and the presence of this dip might be used as a counter-argument for positive flux-period correlation, and the significance of the QPO detection in this period (between $\sim 1.5 \times 10^4$ and  $\sim 1.7 \times 10^4$ s) is relatively high (between 90 and 95\% confidence level). However, in most of the Segment 1 data the significance of the QPO detection is below 90\%, and in addition the dynamical range of the measured quantities (including the dip period) is lower than in the Segment 2 while the dispersion is large. As a result, there is no correlation between the flux and period in this segment alone. If all Segment 1 data points are plotted in Fig. 5, they  group in the long period part of the diagram without showing an expected rise in the flux. The dip partially shows an extreme trend, with the three data points having the lowest value of the flux and the highest value of the period, but all other data points during the dip (18) overlap with the rest of the distribution, so the trend seen during this dip cannot be considered as significant. Thus, we concentrate on the result from the Segment 2 alone in our discussion below.

Since the QPO of a 3733 s period smears the correlation at the longer
time-scales, we extract the short-time trend from the light curve by applying
the
moving average approach, with a 5000 s window function. In consequence, we
obtain a smoothed curve, free from temporal high-frequency flux variability. 

If we now use this long time-scale component for a whole light curve, the
measured delay between the flux and the period is somewhat shorter, $700$ s,
the correlation coefficient is slightly higher, 0.40, and the slope remains
comparable, 0.38. If  only the Segment 2 data are used, i.e. when the QPO
is the most prominent, the delay is slightly shorter than for a whole
light curve, now $1200$~s, the correlation becomes more significant ($r =
0.80$), and the slope is again close to 1 (see Fig.~\ref{fig:flux_period},
lower panel), although a bit smaller ($0.92 \pm 0.03$). 

The results for the Segment 2 (when the QPO and the correlation are strong) for
the original and the smoothed light curve are shown in
Fig.~\ref{fig:flux_period}, respectively. Both of them show similar trends, but
the dispersion is clearly lower for the smoothed light curve. Straight lines
show the best linear fits. This linear relation with the slope of 1 (i.e., a
simple proportionality of the flux and QPO period with an offset from zero) is
an important constraint for the models of QPO itself. 

Because the amplitude of the wavelet map also varies considerably, we 
tested for other possible correlations as well. The weak correlation between the wavelet 
amplitude and the QPO period was present but it was less significant than the 
correlation of X-ray flux vs. QPO period. The highest value of the correlation 
coefficient for the amplitude-period was 0.47 (lower than 0.52 for the 
flux-period correlation
for un-smoothed flux and 0.80 for a smoothed flux; all correlations measured for 
Segment 2 only), and the delay was again 1400 s. The X-ray flux, whether 
smoothed or not, was uncorrelated with the wavelet amplitude.

\section{Discussion}

We performed the wavelet analysis of the XMM X-ray lightcurve of REJ 1034+396
from 2007~May~31, when the source showed a clear QPO phenomenon (G08). Other
observations of this source did not show such a signal on the top of 
the typical red noise (C. Done,
private communication)
but this is not surprising in view of the low duty cycle for a QPO phenomenon 
in the Galactic black holes. The phenomenon is most likely to be an example of the
high-frequency QPO, as discussed by Middleton et al. (2009) and Bian (2009).

Our analysis of the X-ray light curve of RE J1034+396 through wavelet
technique indicates the dependence of the QPO peak position on time,
and the change of the QPO period correlates with the X-ray flux.
 This correlation 
implies a common source of perturbation. 

Qualitative study of this relation suggests a linear 
dependence between the
flux and the QPO period. Because the trend in the QPO period is likely to 
be connected with the change in the size/shape of the oscillating region, we
can try to constrain the geometry by estimating the rate of change of both
the frequency and flux.


In the simplest dimensional analysis approach, we can assume that the local properties of the
emitting 
region do not depend on the region size. The characteristic period is then 
likely
to be proportional to the total size of oscillation region, while the 
emitted flux is expected to rise with a region volume. Therefore, within this
simplest scenario we would expect $flux \propto R^3$ for a spherical region, 
$flux \propto R^2$ for a flat region and  $flux \propto R$ for a linear
emitting region. Only the last one would reproduce the observed linear relation
between the flux and the oscillation period.

\subsection{Oscillating and dissipating tori}

The more realistic models of the QPOs are based on the computations of the tori
around black holes. The proper frequencies of these tori oscillations are on the
order of the Keplerian period, $P$, at the torus outer edge, $R$. Therefore, in
this case $P \propto R^{3/2}$. The overall torus emissivity is likely to be
related to the accretion dissipation inside this region. In a simple accretion
disk approach, the dissipated energy depends mainly on the inner radius,
$R_{in}$, and only weakly on the outer radius $R$ unless these two radii are
very close. For the latter, the dissipation inside is roughly proportional
to $(R-R_{in})$. Therefore, for a moderately wide torus the significant change
of the flux following the change of the period is not expected, and for a narrow
ring the flux response may be arbitrarily strong. However, a linear response
does not seem to be favored.

The quantitative models of the QPOs published so far do not show the connection
between the average flux and the adopted outer radius required in testing the
QPO mechanism according to our finding (see e.g. Schnittman \& Rezzola 2006;
Horak 2008).

\subsection{Emission of the magnetic flare on the Keplerian orbit}

Typical AGN light curves are stochastic but, occasionally, the distinct
large flares have been reported either in the light curve itself (Ponti
et al. 2004) or in the periodic behavior of the iron line (e.g. Iwasawa
et al. 2004; Turner et al. 2002, 2006; Guainazzi 2003; Yaqoob et al. 2003,
Dov\v{c}iak et al. 2004; Markowitz et al. 2006). A magnetic loop can expand 
or contract in length, eventually producing the linear flux$-$period relation
required by the data. 

However, in the case of the disk, this flare is likely to perform an orbital
motion around a central black hole, sharing the bulk motion with the
underlying accretion disk. Then the relativistic effects will also
couple the flux and the period, but in the opposite direction; the Doppler boosting
enhances both the flux and the frequency, so the anti-correlation between the 
flux and the period is expected, in contrast to what is observed. Only if
the flare is very distant, or observed at a very low inclination angle, the
relativistic effects will be negligible. Here, we study the required conditions
with a simple model. 

We assume that the intrinsic flare oscillation has a fixed frequency,
$f_0$, and the flare is anchored in a Keplerian accretion disk. The
effect of relativistic periodic change of the flare frequency and 
luminosity can be modeled by parametrizing the flare position radius
$r$, measured in gravitational radius units ($R_g = GM/c^2$), 
the initial phase $\phi_0$, and the inclination of an observer, $i$
with respect to the accretion disk. We approximate the flare geometry by
a point-like source radiating isotropically. In the center, we assume a
Schwarzschild black hole because the probable radii are large and at
these distances the corrections because of the black hole spin are
negligible (Murphy et al. 2009).

The flare frequency as seen by the observer can be calculated from the
formula $f(t) = f_0 g(r,\phi)$,
where the $g(r,\phi)$ factor is (Pech\'a\v{c}ek et al. 2006)
\begin{equation}
g(r,\phi) =  \frac{r^{1/2}(r-3)^{1/2}}{ r + 
\sin\phi\; \sin i\; \surd[r - 2 +4(1 + \cos \phi \sin i )^{-1}]} \ .
\end{equation}
The flare in Keplerian orbital motion has the position angle $\phi$ varying with
time. However, the observed rate of $\phi(t)$ is not uniform because of the
time delays owing to the light-travel time that photons experience from different
points in the orbit. The time delay can be approximated as follows
\begin{equation}
t = {\phi \over \Omega_{\rm K}(r)} + (r - 1) (1 - \cos \phi) \sin i + 
2 \ln { 1 + \sin i \over 1 + \sin i \cos \phi},
\label{tpech}
\end{equation}
where $\Omega_{\rm K}(r)$ is the Keplerian angular velocity. 
The flare intensity is then
\begin{equation}
L = L_0\; g(r,\phi)^4 \left[1 + 
{1 \over r}{1 - \sin i\; \cos \phi \over 1 + \sin i\; \cos \phi}\right]\; \cos i
\ .
\label{eq:light curve}
\end{equation}

\begin{figure}
\includegraphics[angle=0,width=0.475\textwidth]{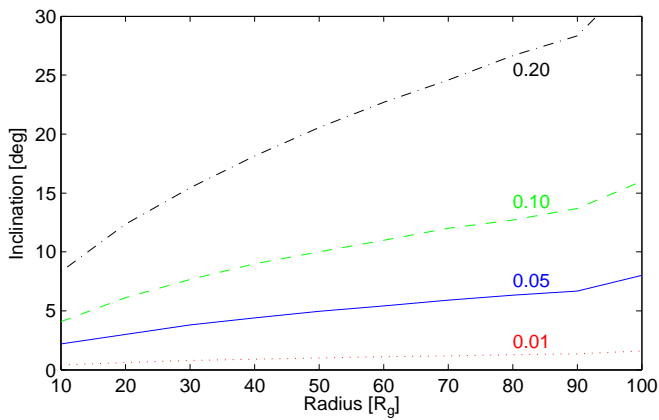}
\caption{Expected fractional flux enhancement anti-correlated with the QPO
period for a 
flare in the Keplerian motion as a function of the orbital radius and the
inclination angle. The flare must be located far away or the inclination must be low
to ensure that this effect does not suppress the observed opposite trend.
Amplitudes larger than $\sim 0.05$ (above the continuous blue line) are highly 
improbable, and low inclination 
(below 10 deg) is required
for a flare model.}
\label{fig:incli_radius}
\end{figure}

Assuming a black hole mass $2 \times10^6 M_{\odot}$ we determined the 
fractional amplitude of a flare owing to the relativistic motion for a range of
radii and inclination angles. Because the QPO itself contains about 10\% 
of the total power of the source (Gierlinski et al. 2008), 
and a fractional change in the observed trend
is about 5\% (see Fig.~\ref{fig:folded_curve}), then a relativistic effect 
at the level of 20\% would totally cancel the observed opposite trend and only
the effect at the level about 5\% may pass unnoticed. We show the lines of a constant 
fractional flux enhancement as functions of the flare location 
in Fig.~\ref{fig:incli_radius}.  The blue continuous line in this plot marks acceptable
range, so the disk inclination should be lower than 2 - 3 deg for flares close 
to a black hole , and
inclinations up to 7-8 deg are acceptable for flares at a distance of 100 $R_g$.
We have no observational constraints for RE J1034+396, but the inclinations of other 
radio quiet AGN are typically estimated to be about 30 deg. 

Very distant flares, at radii much larger than 100 $R_g$ are not excluded by this consideration. Then the relativistic effects become unimportant and the flare luminosity 
changes only because of its intrinsic evolution which may involve pulsations.

\subsection{Shocks and spiral waves}

In a series of papers (see e.g. Das et al. 2003, Mondal et al. 2009) 
the QPO phenomenon was modeled as caused by the oscillations of 
a shock that forms in a thin accretion flow. The applicability of the model is
not immediate because 
RE~J1034+396 is a disk-dominated source. However, a coronal
flow may be present with properties similar to the predicted in the model
without the underlying Keplerian disk. In those models, there is a relation
between the outflow rate (which may be used as a proxy to luminosity) and the
frequency, and this relation for low values of the entropy an angular momentum
has a power-law shape, but an index of only 0.25 (Das et al. 2003, Fig.~3, upper
left panel, two lowest curves). However, a better proxy for the system luminosity
yields an index consistent with the observed one (Das et al., in preparation).  

Spiral waves were also suggested as a possible explanation of the QPO
phenomenon,
(e.g. Tagger \& Varniere 2006; see also Chan et al. 2009 for a study in the 
context of Sgr A*) and because of their predominantly linear structure they may
possibly give the
requested flux-QPO period relation. The spiral wave rotates with constant
angular velocity, and the average rotation speed 
of this pattern is likely to be considerably lower than Keplerian in the
innermost 
part of the disk, although the matter itself is still roughly in Keplerian
motion. It may reduce the strength of the relativistic effects and act against
the positive correlation between the flux and frequency resulting from a slow
pattern evolution. However, the quantitative modeling would be needed to see if
the coherent model of the phenomenon can be found within the frame of this
scenario. Simulations of the flare-like events for Sgr~A* perhaps show
occasional traces of the correlated change in the period and flux for a given
oscillation (see Fig.~12 of Chan et al. 2009), but the results may not directly
apply to REJ~1034+396. 

\section{Conclusions}

The wavelet analysis of the X-ray light curve of RE~J1034+396 allowed us to see
a
trend in the QPO period during a single QPO event. An increase in the QPO
central period seems to be accompanied by a proportional increase in the
X-ray flux.

This observation is unique because the source did not show any QPO phenomenon at another 
time apart from the observation analyzed by Gierlinski et al. (2008), and the claimed QPO
events in other AGN are less significant (Espaillat et al. 2008; Lachowicz et
al. 2009). The detection of the trend within the QPO is difficult, but the signal
seems to be there. If the result indicating the proportionality between
the flux and the QPO period during a single QPO event is a general property,
then it implies very strong constraints on the QPO mechanism. The oscillating
tori seem less likely, the linear structures are favored (shocks or spiral
waves), and they should not be involved in Keplerian motion around a black
hole if they are localized at a small radius, unless the viewing angle in
REJ~1034+396 is exceptionally low. The models of these structures have to be
further developed and they should, at some point, also explain why
the oscillating hard X-ray emitting plasma responsible for QPOs (Middleton et
al. 2009; van der Klis 2006; \.Zycki et al. 2007) has to coexist with
the accretion disk.

In galactic sources single QPO high-frequency events are unresolved, therefore
this newly found correlation cannot be tested. However, future X-ray
observations (e.g. with IXO) will bring many more light curves of AGN
with a quality good enough to perform a similar analysis, even if the QPO duty
cycle is relatively low, as in galactic black holes.

\acknowledgements
Wavelet software was provided by C. Torrence and G. Compo, and is available at 
the URL: http://atoc.colorado
.edu/research/wavelets/.
We thank Piotr \.Zycki and Alex Schwarzenberg-Czerny for very helpful discussions. 
This work was supported in
part by grant NN 203 380136. VK and MD acknowledge the Czech Science Foundation
grants 205/07/0052 and 202/09/0772.

\end{document}